# Spectroscopic Fingerprint of Chiral Majorana Modes at the Edge of a Quantum Anomalous Hall Insulator / Superconductor Heterostructure

J. Shen[1], J. Lyu[1,2], J. Z. Gao[1], Y.-M. Xie[1], C.-Z. Chen[1], C.-w. Cho[1], O. Atanov[1], Z. J. Chen[3,4], K. Liu[3,4], Y. J. Hu[5], K. Y. Yip[5], S. K. Goh[5], Q. L. He[6,7], L. Pan[6], K. L. Wang[6,*], K. T. Law[1,*] and R. Lortz[1,*]

[1]*Department of Physics, The Hong Kong University of Science and Technology, Clear Water Bay, Kowloon, Hong Kong;*
[2]*Department of Physics, Southern University of Science and Technology, 1088 Xueyuan Road, Nanshan District, Shenzhen, Guangdong Province, China;*
[3]*Physics Department, University of California, Davis, CA 95616, USA;*
[4]*Physics Department, Georgetown University, Washington, DC 20057, USA;*
[5]*Department of Physics, The Chinese University of Hong Kong, Shatin, New Territories, Hong Kong;*
[6]*Department of Electrical and Computer Engineering, Department of Physics, and Department of Materials Science and Engineering, University of California, Los Angeles, CA 90095, USA;*
[7]*International Center for Quantum Materials, School of Physics, Peking University, Beijing, 100871, China.*

*Corresponding authors:*
Rolf Lortz (lortz@ust.hk)
Department of Physics, The Hong Kong University of Science and Technology, Clear Water Bay, Kowloon, Hong Kong,
(852) 2358-7491

Kam Tuen Law (phlaw@ust.hk)
Department of Physics, The Hong Kong University of Science and Technology, Clear Water Bay, Kowloon, Hong Kong,
(852) 2358-7970

Kang Wang (wang@ee.ucla.edu)
Department of Electrical and Computer Engineering, Department of Physics, and Department of Materials Science and Engineering, University of California, Los Angeles, CA 90095, USA,
(310) 825-1609

**Abstract**
With the recent discovery of the quantum anomalous Hall insulator (QAHI), which exhibits the conductive quantum Hall edge states without external magnetic field, it becomes possible to create a novel topological superconductor (SC) by introducing superconductivity into these edge states. In this case, two distinct topological superconducting phases with one or two chiral Majorana edge modes were theoretically predicted, characterized by Chern numbers ($\mathcal{N}$) of 1 and 2, respectively. We present spectroscopic evidence from Andreev reflection experiments for the presence of chiral Majorana modes in a Nb / $(Cr_{0.12}Bi_{0.26}Sb_{0.62})_2Te_3$ heterostructure with distinct signatures attributed to two different topological superconducting phases. The results are in qualitatively good agreement with the theoretical predictions.

**Introduction**
Topological SCs which host Majorana zero energy modes [1] have attracted great interest [2-5], since they represent ideal candidates for fault-tolerant qubits with non-abelian exchange statistics [6-8]. One way to realize Majorana states motivated by a model proposed by Kitaev [9] is based on semiconductor nanowires with strong Rashba spin-orbit coupling in which superconductivity is induced by the proximity to an s-wave SC [10-12]. This forms topological superconductivity of Class D, which breaks the time reversal symmetry. Majorana zero modes at each end of the wire were identified recently by the characteristic quantized zero bias conductance peak in their tunneling characteristics [13,14]. An alternative system is the chiral p-wave superconductor, which has been reported to occur intrinsically in $Sr_2RuO_4$ [15] and predicted to occur when a QAHI with chiral edge modes is driven superconducting via the proximity effect [16-19]. A topological SC of D class is then expected, which has two distinct topological superconducting phases with either two ($\mathcal{N} = \pm2$) or one ($\mathcal{N} = \pm1$) chiral Majorana edge states. In the QAHI, dissipationless conductive chiral quantum Hall edge states classified by a Chern number $C = \pm 1$ are generated by introducing perpendicular ferromagnetism into topological insulator (TI) films by doping with magnetic ions [20,21]. During a magnetization reversal an additional topologically trivial insulating phase was observed [22,23]. When superconductivity is induced into these chiral edge channels, a topological superconducting phase is expected, which is classified by the Chern number $\mathcal{N}= \pm2$. The chiral QAHI edge state then consists of two degenerated chiral Majorana states [16-19]. When approaching the trivial insulating phase, one of the Majorana modes diffuses into the bulk and annihilates, forming an intermediate phase $\mathcal{N} = \pm1$ with only one chiral Majorana state as a distinct topological superconducting phase [16-19]. These chiral Majorana edge states would be excellent candidates for the realization of scalable topological quantum computation by constructing networks of quasi-one-dimensional QAHI / SC structures [24]. In Ref. [25] a device was presented in which a superconducting strip was deposited on a QAHI. Quantized half-integer conductance plateaus with a value $0.5e^2/h$ were reported and interpreted as a consequence of the $\mathcal{N} = \pm1$ phase [26,27]. However, these results provoked various controversies and it was theoretically argued [28,29] and experimentally demonstrated [30] that such plateaus could occur under certain circumstances without superconductivity.

In this letter, we present an alternative approach based on Andreev reflection experiments using nanoscale point contacts at the edge of a QAHI / SC heterostructure, to investigate the low-energy excitations during the topological transitions. We find a dip-like structure surrounded by two peaks, which agrees with the theoretical prediction of the state $\mathcal{N} = \pm2$ [19], in which the resonant Andreev reflection signals of the two Majorana modes cancel each other at small bias voltage [19]. As the field is increased and a reversal of magnetization is induced, the dip transforms into a plateau as expected when the state $\mathcal{N} = \pm1$ is formed [19]. These results suggest the presence of two distinct topological superconducting phases.

**Results**

A sketch of this edge point contact device is shown in Fig. 1(a). The heterostructure of total area 5 x 2 mm$^2$ consisted of a 6 nm $(Cr_{0.12}Bi_{0.26}Sb_{0.62})_2Te_3$ film on a GaAs (111) substrate and was entirely covered by a 200 nm Nb layer, with nano-point contacts in the form of 200 nm up to 500 nm wide metal strips at the edge of the heterostructure. Our nano-point contact method is a very local spectroscopic probe and thus largely unaffected by domain structures [27-29] in the QAHI film. The contacts are so small that edge states of individual domains can be investigated [30]. For more details on sample growth and device fabrication, see the Methods section of this article. Fig. 1(b) shows the sharp resistive $T_c = 7.2$ K of Nb and Fig. 1(c) shows the upper critical field transition in the magneto-resistance at $H_{c2}$(273 mK) = 9 T, demonstrating a high quality of the Nb film.

In our nano-point contact setup it was inevitable and even essential that a parallel contact with the Nb layer was established and the measured d$I$/d$V$ spectra typically contained contributions from both layers. Different contacts showed different characteristics due to different barrier heights between electrode and Nb or $(Cr_{0.12}Bi_{0.26}Sb_{0.62})_2Te_3$ layers. In Fig. 1(d) we show a selection of typical raw data of Contact 1, which was in the highly transparent Andreev limit (contact resistance < 100 Ohm). The d$I$/d$V$ data are dominated by a broad peak centered at zero bias. The main contribution can be identified as a result of Andreev reflections from the Nb layer, as the fits with the BTK model [31] show ($Z$=0.01, $\Delta$=1.6 meV, $\Gamma$ =0.6 meV), where $Z$ is the effective barrier strength, $\Delta$ the superconducting gap value and $\Gamma$ the finite quasi-particle lifetime or broadening parameter. Note that these parameters are dominated by the contact to Nb and the parameters for the parallel contact to the QAHI may differ somewhat. For more details on fitting, see the Methods section. Note that due to the high $H_{c2}$ of Nb, no significant magnetic field dependence is expected for the small field range of interest in our experiments. The Nb contribution in zero field is superimposed by a small feature with two symmetric peaks that flatten the tip of the Nb delta shape. To isolate the contribution of proximity induced superconductivity in the QAHI, we apply a 175 mT magnetic field known to drive the QAHI into the trivial insulating state [22,23,25,32] in which the conducting edge state vanishes and thus proximity-induced superconductivity is entirely suppressed. At a sufficiently low bias voltage (below ±2 mV), these spectra originate only from the Nb contribution, which were then used to isolate the QAHI contribution in data in lower and higher fields. At higher bias voltages strong oscillations are seen, which are attributed to bulk electronic bands in the QAHI.

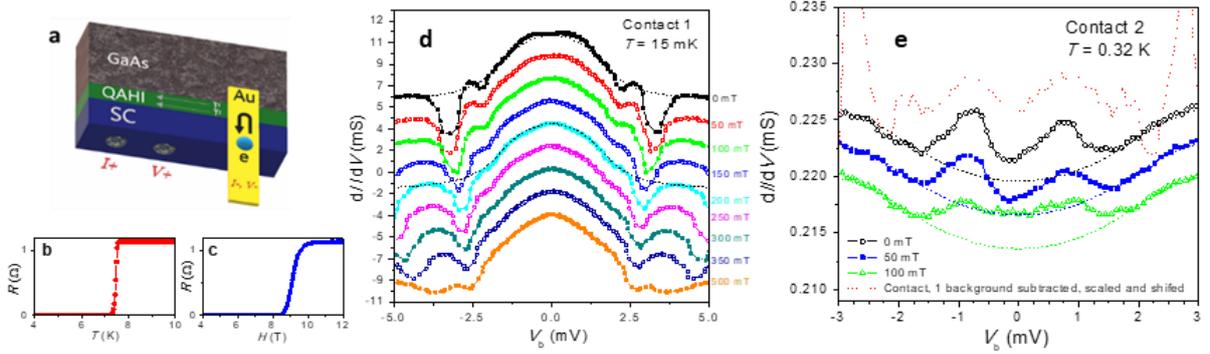

FIG. 1. (a) Sketch of the point contact device on the edge of a QAHI / SC heterostructure. It features normal contacts on the surface of a superconducting Nb layer and point contacts realized by thin Au strips crossing the edge of the heterostructure. The arrows illustrate the Andreev reflection process. Note that the dimensions of the sketch are not to scale: The actual film dimensions were 5 mm x 2 mm, while the thickness of the QAHI was 6 nm and the Nb thickness was 200 nm. (b) Zero-field electrical resistance measured by 4 ohmic contacts on the surface of the Nb layer. The sharp resistance jump defines $T_c$ = 7.2 K. (c) Magnetoresistance measured at $T$ = 273 mK as a function of a magnetic field applied perpendicular to the film. The jump marks $H_{c2}$ = 9 T. (d) High-transparency point contact spectroscopy raw data d$I$/d$V$ of Contact 1. A dominant superconducting Nb contribution is seen as a large positive Andreev reflection peak [31]. (e) Low-transparency point contact spectroscopy data d$I$/d$V$ of another QAHI / SC heterostructure (Contact 2). The Nb contribution shows up as a broad shallow gap. The QAHI contribution can be seen here without any background subtraction. Scaled zero-field data of Contact 1 after subtraction of the Nb contribution has been added for comparison. Note that the additional oscillations at higher bias voltage beyond ±2 mV are attributed to the bulk bands of the QAHI. Fits of the SC Nb background with the BTK model [31] have been included in (d) and (e) (dashed lines).

In Fig. 1(e) we show data from a contact in the low transparency tunneling limit (Contact 2) fabricated on the same heterostructure. Here, the Nb contribution appears as a shallow, broad gap. A fit of this contribution with the BTK model is included ($Z$=8, $\Delta$=2.3 meV, $\Gamma$=1.5 meV), which describes the data well with a gap value of 2.3 meV, which corresponds to the literature data of bulk Nb [33]. Two peaks surrounding a dip appear here at lower bias voltage, which transform in a plateau in a field of 100 mT, shortly before the QAHI enters the trivial insulating state. As we will show below, this is the expected signature of chiral Majorana fermions in the QAHI layer. While this contact would be ideal to investigate these features without background subtraction, it unfortunately became unstable and therefore we will focus on Contact 1, which has the same features attributed to a superconducting QAHI edge state after separation of the Nb background. For comparison, we have included the 0T data of Contact 1, which, in the region of the proximity gap at low bias voltage, shows excellent agreement with the raw data of Contact 2. This demonstrates that the 175 mT data represents an ideal background for all data of Contact 1, which is justified since no edge conduction occurs in the trivial insulating state and thus the QAHI does not contribute to the point contact spectrum in this field. For chiral Majorana modes, it has been theoretically shown that any form of tunneling from a lead to the edge of the superconducting QAHI occurs via resonant Andreev reflection [19]. Therefore, although Contact 1 & 2 have completely different transparency, the presence of Majorana modes is expected to manifest in form of the same shape in d$I$/d$V$, while less transparent contacts only reduce the amplitude.

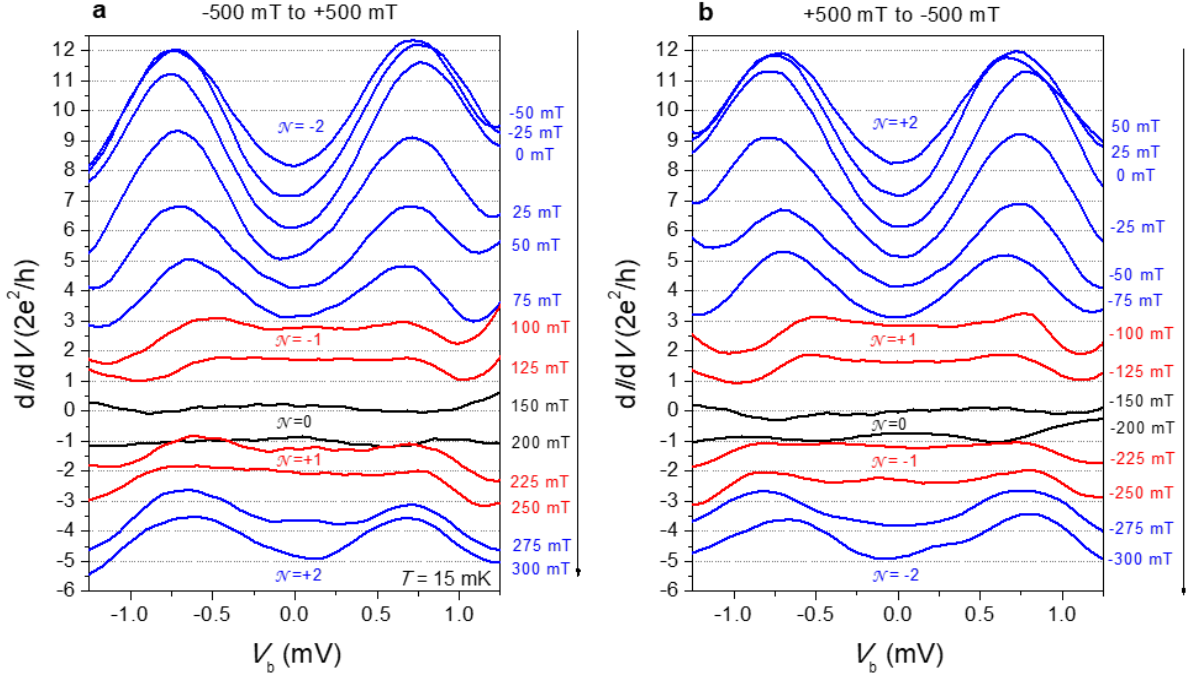

FIG. 2. Differential conductance of a point contact at the edge of the $(Cr_{0.12}Bi_{0.26}Sb_{0.62})_2Te_3$ / Nb heterostructure at 15 mK (Contact 1), after separation of the Nb contribution. The data in (a) was taken starting from a negative field upon increasing the field in 25 mT increments to positive fields, inducing the QAHI's magnetization reversal at ~175 mT, while for the data in (b) the opposite direction of field change was used to study the entire hysteresis range of magnetization. Dips surrounded by two peaks mark the $\mathcal{N} = \pm 2$ topological superconducting regions, while flat plateaus are observed at +100 mT and +125 mT (a), or -100 mT and -125 mT (b), which are attributed to the $\mathcal{N} = \pm 1$ topological superconducting phases. The latter occurs shortly before the QAHI enters the trivial insulating region ($\mathcal{N}=0$). When the absolute value of the magnetic field increases further, a re-entry of the $\mathcal{N} = \pm 1$ phase in the form of the plateau at 225 mT (-225 mT) is seen, and finally the characteristic dip of the $\mathcal{N} = \pm 2$ phase occurs again. These features are unique to $(Cr_{0.12}Bi_{0.26}Sb_{0.62})_2Te_3$ in proximity to the superconducting Nb layer and occur within the insulating bulk gap region of $(Cr_{0.12}Bi_{0.26}Sb_{0.62})_2Te_3$. Offsets of $2e^2/h$ have been added for better clarity, except for the ±150 mT data.

In the following we investigate with Contact 1 how the contribution to $dI/dV$ from the QAHI changes when the QAHI is driven back and forth through the magnetization reversal, and thus investigate the entire range of magnetization hysteresis to search for signatures of topological transitions in a superconducting QAHI state. In Fig. 2, we show two series of differential conductance data from Contact 1 after subtraction of the Nb contribution, which were taken during the magnetization reversal of $(Cr_{0.12}Bi_{0.26}Sb_{0.62})_2Te_3$ by repeated measurements upon small 25 mT increments (decrements) of the magnetic field, starting from a negative (positive) field of -500 mT (+500 mT) towards a positive (negative) field of +500 mT (-500 mT). The start field of -500 mT (a) or +500 mT (b) is used to fully magnetize the QAHI insulator before the experiment. In Fig 2(a) the field is changed from negative to positive fields and in Fig 2(b) in the opposite direction. Due to the magnetic hysteresis, magnetization is maintained when the field is reduced to zero and the reversal of magnetization is induced when the field reaches the opposite polarity in the field range between 150 mT (-150 mT) and 200 mT (-200 mT) [22]. The data have been shifted vertically by $2e^2/h$ (except for the ±150 mT data) for better visualization. Both datasets show exactly the same trend. At -100 mT (+100 mT), the data show a characteristic dip at zero bias framed by two peaks around ±0.75 mV, indicating the presence of proximity-induced superconductivity in the QAHI. As we will show later, we attribute them to the $\mathcal{N} = -2$ ($\mathcal{N} = +2$) topological superconducting states in which the cancellation of Andreev reflection amplitudes of a pair of chiral Majorana modes causes a suppression of the conductance around zero bias [19]. The plus or minus signs refer to the direction of the QAHI magnetization. As the magnitude of the applied field is gradually reduced, the size of the dip and peaks increases and reaches a maximum at zero field. After crossing the zero field the amplitude further decreases until at +100 mT (-100 mT) the dip changes into a plateau with relatively sharp edges at ±0.7 mV, which we will show later is attributed to the $\mathcal{N} = -1$ ($\mathcal{N} = +1$) topological superconducting states. The plateau persists at +125 mT (-125 mT), while it vanishes at +150 mT (-150 T). By further increasing the field strength to +175 mT (-175 mT) and 200 mT (-200mT), the data becomes flat, indicating that the QAHI enters the trivial insulating state in which proximity-induced superconductivity vanishes during the magnetization reversal. Note that the same transition to a plateau can be seen in the raw data of Contact 2 in Fig. 1(e) at 100 mT, without the need for background subtraction.

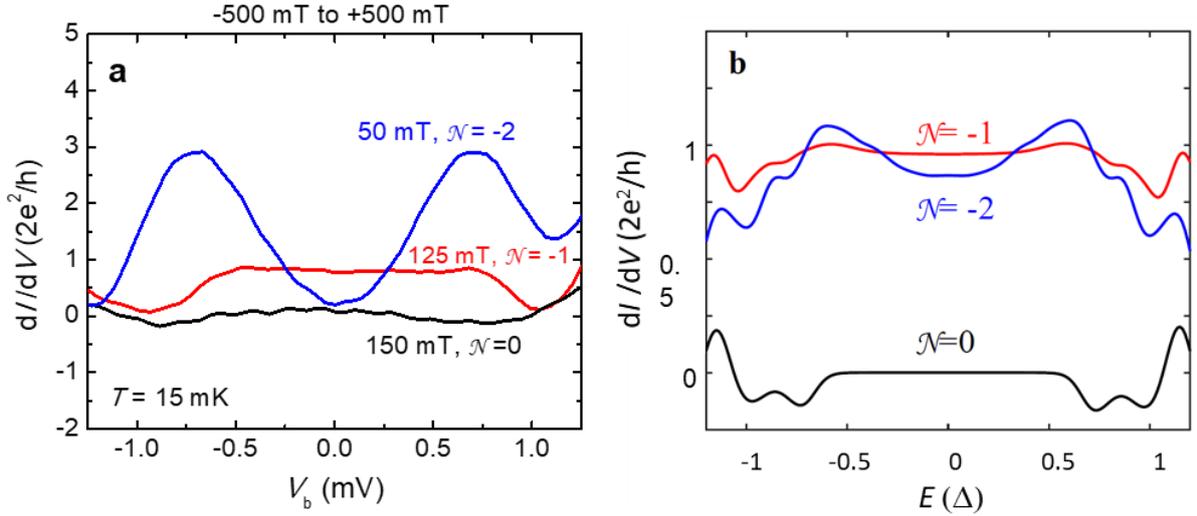

FIG. 3. (a) Selected data of the differential conductance of Contact 1 ($T$ = 15 mK) without offsets to illustrate the difference of the spectra in the different topological regimes. (b) Simulated d$I$/d$V$ as a function of the bias voltage energy (normalized by the SC energy gap $\Delta$) for selected magnetization energy values of $M_z$=0.01$m_0$ ($\mathcal{N} = 0$), -1.2$m_0$ ($\mathcal{N} = -1$), -1.85$m_0$ ($\mathcal{N} = -2$). Dips at zero bias mark the chiral fermionic modes in the $\mathcal{N} = \pm 2$ topological superconducting region, while a flat plateau indicates the single chiral Majorana mode in the $\mathcal{N} = \pm 1$ topological superconducting phase. In contrast, the trivial insulating state with $\mathcal{N} = 0$ shows a vanishing conductance in a range around the zero bias voltage. The details of the simulations can be found in the Supplementary Materials.

In Fig. 3 (a) we show selected data from the three topological regimes without shift to further illustrate the differences in the spectra. In the $\mathcal{N} = 1$ state the plateau-like feature differs significantly from the spectra in the $\mathcal{N} = 2$ state with its characteristic dip around zero bias and from the trivial insulating state in which the conductance remains zero. The plateau of the $\mathcal{N} = 1$ state clearly differs from the expected signature of a trivial metallic state, in which the plateau would extend to a wide voltage range. Here the $\mathcal{N} = 1$ plateau is limited to the bias voltage range associated with the superconducting proximity gap, which is about 0.7 meV, and vanishes abruptly at higher bias voltage. It is also important to note that the drop in d$I$/d$V$ forming the edge of the plateau at 0.7 meV is unlikely to be caused by the superconducting Nb background, since the pairing gap of Nb is estimated to be at least 1.6 meV in our experiment. Furthermore, the height for this contact remains close to the conductance quantum $2e^2/h$. We would also like to point out that at both Contact 1 and Contact 2 a plateau of the same shape is observed, although the height of the tunneling barrier differs strongly. This can be plausibly explained by resonant Andreev reflection processes in the presence of Majorana modes [19], whereas the spectra of ordinary superconductors would look completely different in the two regimes [31].

Upon further increasing (decreasing) the applied field in 25 mT steps the QAHI continuous to evolve from the trivial insulating state towards the fully magnetized $C = +1$ state. At 225 mT (-225 mT) a flat plateau associated with the $\mathcal{N} = +1$ ($\mathcal{N} = -1$) phase reappears. When sweeping the field further, at +275 mT (-275 mT) the plateau changes back into two peak-like structures surrounding a dip around zero bias. This indicates a re-entry of the topological state with two chiral Majorana modes ($\mathcal{N} = +2$). We would like to point out once again that the main features, namely the transition from a peak-dip like feature ($\mathcal{N}= 2$) to a plateau-like feature ($\mathcal{N}= 1$) are also visible in the raw data of Contact 2 without the need to subtract background data, which gives us a high degree of confidence in these results.

Further data of Contact 1 are included in the Supplementary Materials [34]. It shows selected data recorded during an additional field cycle and data at a higher temperature of 280 mK, thus demonstrating the reproducibility of the results. We also show similar data from another heterostructure grown under identical transition (Contact 3 & Contact 4). For our contacts in the high-transparency Andreev limit (Contact 1 & Contact 3) the plateau associated with the $\mathcal{N} = \pm1$ state appears to be not far from quantization at a conductance of $\sim 2e^2/h$.

**Discussion**

In the following, we will show that the transition from a dip-like feature at zero bias to a conductance plateau as a function of magnetic field is indeed expected when the QAHI changes from the $\mathcal{N}= +2$ state with two chiral Majorana edge modes to the $\mathcal{N}= +1$ state in which only one chiral Majorana mode remains at the edge. The magnetic TI thin film $(Cr_{0.12}Bi_{0.26}Sb_{0.62})_2Te_3$ can be described by $\mathcal{H}_{TI}=\sum_k \Psi_k^\dagger H_{TI}\Psi_k$ in the basis of $\Psi_k = [\psi_{t\uparrow},\psi_{t\downarrow},\psi_{b\uparrow},\psi_{b\downarrow}]^T$, where $\psi$ denotes an electron annihilation operator [26]. Explicitly,

$$H_{TI}(\mathbf{k}) = \begin{pmatrix} h_t(\mathbf{k}) & m_\mathbf{k} \\ m_\mathbf{k} & h_b(\mathbf{k}) \end{pmatrix},$$

with $h_{t,b}(k) = \pm v_F(k_y\sigma_x - k_x\sigma_y) + M_z\sigma_z$. Here the Pauli matrices $\sigma_{x,y,z}$ act on spin ($\uparrow/\downarrow$) space and $t/b$ denotes the top/bottom surface of the TI thin film and $m_\mathbf{k} = m_0 + m_1 k^2$ describes the hybridization between the two surfaces, while $v_F$ is the Fermi velocity of the surface Dirac fermions. The magnetization energy, $M_z$, is caused by intrinsic magnetic dopants as well as an external magnetic field. The magnetic TI thin-film Hamiltonian is in $C = M_z/|M_z|$ QAHI phase with a chiral fermionic edge mode at the boundary when $|M_z| > |m_0|$, while becoming a $C = 0$ normal insulator when $|M_z| < |m_0|$.

In proximity to an *s*-wave SC, the effective Hamiltonian of the $(Cr_{0.12}Bi_{0.26}Sb_{0.62})_2Te_3$/SC heterostructure in the Bogoliubov-de-Gennes (BdG) space becomes

$$\mathcal{H}_{BdG} = \begin{pmatrix} H_{TI}(\mathbf{k}) - \mu & \Delta_\mathbf{k} \\ \Delta_\mathbf{k}^\dagger & -H^*_{TI}(-\mathbf{k}) + \mu \end{pmatrix}$$

where

$$\Delta_\mathbf{k} = \begin{pmatrix} \Delta_t i\sigma_y & 0 \\ 0 & \Delta_b i\sigma_y \end{pmatrix}$$

Here, $\mu$ is the chemical potential and $\Delta_\mathbf{k}$ denotes the induced *s*-wave pairing with $\Delta_t = \Delta$ and $\Delta_b = 0$. In general, the $\mathcal{N}= \pm2$ superconducting phase is topologically equivalent to the $C = \pm1$ QAHI phase, i.e., a single chiral fermion edge state can be regarded as two branches of chiral Majorana edge states. The $\mathcal{N}= \pm1$ phase is a new topological phase that supports one single Majorana fermion that propagates along the edges of the sample [16-19]. Further details about the Green's Function formalism used can be found in Ref. [19] and in the Supplementary Materials.

Fig. 4 (a) shows the energy spectrum at the Γ point of the topological superconductor, i.e., the proximity coupled QAHI strip with the pairing gap interaction, as a function of $M_z$, normalized by the intrinsic magnetization energy $m_0$. The topological invariant $\mathcal{N}$ can only change at the points where the bulk gap closes [16]. This defines five topologically distinct regions separated by four gap closure points. In the following we will show that the topological regions with one ($\mathcal{N}= \pm1$) or two ($\mathcal{N}= \pm2$) pairs of Majorana edge modes have very different transport properties. A selection of simulated tunneling conductance $dI/dV= \frac{2e^2}{h}Tr(R_{he}R_{he}^\dagger)$ [19] for different magnetization energy values of $M_z$ from a metal lead to the edge of a QAHI/SC heterostructure has been included into Fig 3(b) for qualitative comparison with the experimental data in Fig. 3(a). Here $R_{he}$ stands for the local Andreev reflection amplitude. In the experiment $M_z$ is controlled by the applied magnetic field. In the same way as in the experiment, we start with a negative $M_z$ then increase $M_z$ until the QAHI magnetization reversal occurs. At each gap closure point in Fig. 4, $\mathcal{N}$ changes, first from the $\mathcal{N}= -2$ state via $\mathcal{N}= -1$ to the trivial insulating state. A further increase of $M_z$ triggers a re-entry of the superconductivity via the state $\mathcal{N}=1$ back into the state $\mathcal{N}= 2$ once the chiral edge states re-appear after the magnetization reversal [22]. During this sequence of $M_z$ values, it is remarkable to note that in the central region around the zero bias both the dip-like feature at zero bias ($\mathcal{N}= 2$) and the conductance plateau ($\mathcal{N}= 1$) are reproduced, in a qualitatively good agreement with the experimental data in Fig. 2.

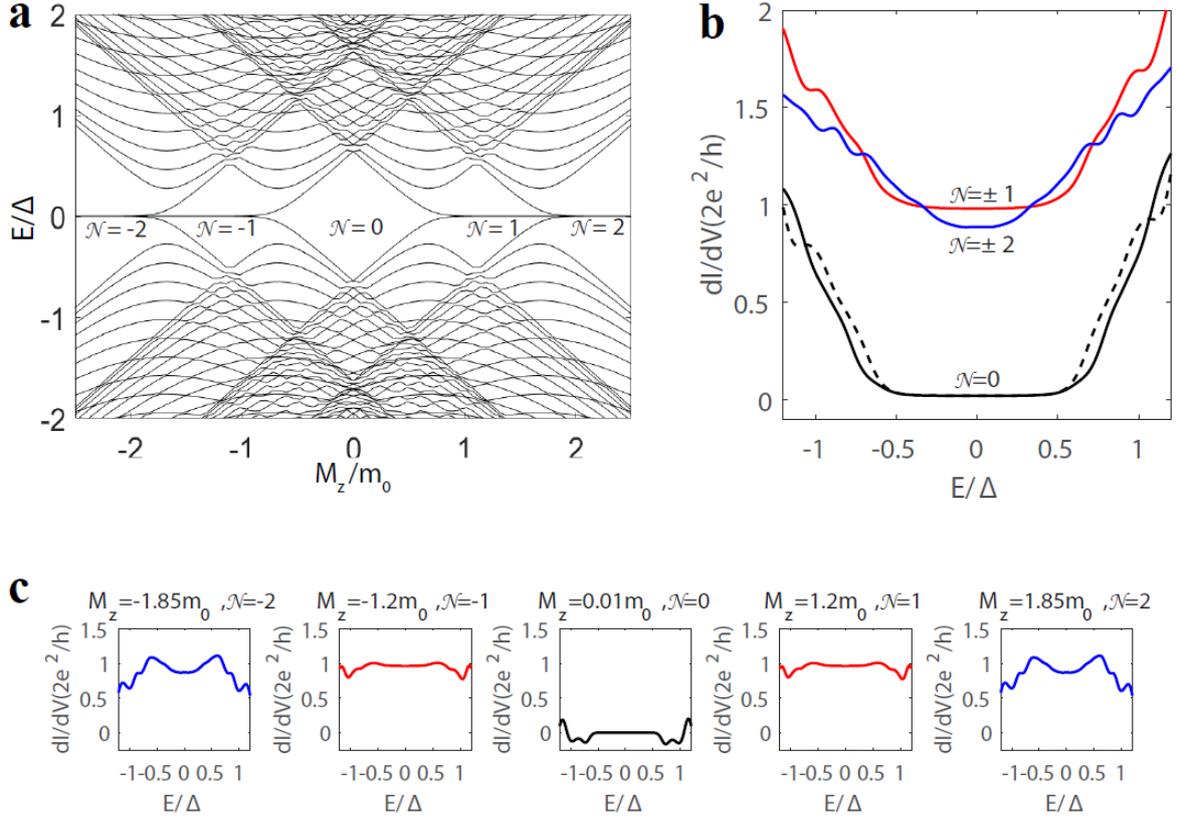

FIG. 4. (a) Energy spectrum at the $\Gamma$ point as a function of $M_z$. The five topologically distinct regions are separated by points at which the bulk gap closes. It shows a series of phase transitions from the $\mathcal{N}= -2$ to the $\mathcal{N}= +2$ topological SC as a function of $M_z$ ($m_0$ is the intrinsic magnetization energy). (b) Simulated tunneling conductance versus energy for the different topological phases. It can be seen that the $\mathcal{N}= 1$ phase exhibits a flat plateau. In contrast, the $\mathcal{N}= 2$ state shows a dip-like shape. These features agree with the results given in Refs. [19] and [36]. The magnetization energies $M_z$ are $-1.85m_0$, $-1.2m_0$, $0m_0$, $0.01m_0$, $1.2m_0$, $1.85m_0$ for $\mathcal{N}= -2$, -1, 0 (dashed line), 0 (solid line), 1, 2 respectively, as listed in Fig. 4(c). (c) Simulated tunneling spectra after background subtraction. The subtracted background is the dashed $\mathcal{N} = 0$ line in (b). After subtraction, the simulation results are in qualitatively good agreement with the experimental data in Fig. 2 and 3a.

In the phase $\mathcal{N}= \pm2$, the Andreev reflection amplitudes of the two branches of the Majorana edge modes cancel each other, which causes a zero bias conductance dip [19]. In the $\mathcal{N}= \pm1$ phase, however, the single branch of Majorana mode induces resonant Andreev reflections within the pairing gap energy. This results in a conductance plateau with a value which is in the order of $2e^2/h$. Both are observed in our experimental data and in the simulated conductance: After magnetizing the sample with a negative $M_z$, we see the characteristic dip around zero bias of the $\mathcal{N}= -2$ state in Fig.3(b). When crossing the boundary to the $\mathcal{N}= -1$ regime at $M_z= -1.8m_0$ in Fig. 4 (a), the dip transforms into a plateau ($M_z= -1.2$ $m_0$). Then the conductance is gradually reduced and approaches zero when the next gap closure occurs at $M_z= -1$ $m_0$. Here, $\mathcal{N}$ changes to 0 and the edge states disappear. In this region of trivial topology with $\mathcal{N} = 0$, the conductance remains at zero for $M_z= 0$ $m_0$. At $M_z= 1.0$ $m_0$ the next gap closure induces the transition into the state $\mathcal{N}= +1$, which triggers a re-entry of superconductivity with one chiral Majorana edge mode. Accordingly, the conductance plateau forms again ($M_z= 1.2$ $m_0$) until at the final gap closure at $M_z= 1.8$ $m_0$, for which $\mathcal{N}$ changes to +2 and the two chiral Majorana modes causing the conductance dips are restored. The experimentally observed conductance plateau appears when the magnetic field after the switching of the sign of the magnetic field is about 100 mT, and then repeats itself near 225 mT, in good qualitative agreement with the simulations. Therefore, we interpret the differential conductance dips and the differential conductance plateaus as signatures of two distinct topological superconducting phases.

While the experimental conductance of the $\mathcal{N} = \pm 1$ plateau suggests a quantization near $2e^2/h$, the simulations show that the absolute height varies slightly with subtle parameter changes (e.g. the tunneling barrier height, see the supplementary Section 5 for a more detail discussion). Precise quantization is only expected for metallic leads coupled to a single Majorana zero mode, but not for leads coupled to dispersive chiral Majorana edge modes. The value of the height of the $\mathcal{N} = \pm 1$ plateau of $2e^2/h$ is therefore likely a coincidence.

**Methods**

The QAHI films $(Cr_{0.12}Bi_{0.26}Sb_{0.62})_2Te_3$ (6 nm thickness) were grown on GaAs (111) substrates in an ultra-high vacuum molecular beam epitaxy system (MBE) at UCLA. The Nb layer (200 nm thickness) was sputtered onto the QAHI layer, which for technical reasons was done at UC Davis. Note that this is different from other approaches [32] where the Nb layer was deposited in the same vacuum chamber and may result in differences in coupling between the two materials. All experiments were conducted by the HKUST group.

Slabs of ~2 mm x 5 mm were cut from the wafer and the long edge of the heterostructure was finely polished before layers of Ti (5 nm) and Au (80 nm) were deposited. A focused ion beam was used to define small isolated metal strips of width (*W*) of a few 100 nm that traverse the entire substrate and film edge. These edge contacts (area: $W \times 206$ nm) represented small nano-point contacts of a few tens of Ohms up to several kOhm resistance at ambient temperature and thus enable point contact measurements from the highly transparent Andreev regime to the low-transparent tunneling regime [31]. It should be noted that a certain roughness of the edge was desired to prevent the superconducting signal from Nb being too dominant. This is the reason for a quite broad gap signature of the Nb signal. Differential conductance vs. bias voltage measurements were conducted in a dilution refrigerator with the magnetic field strictly perpendicular to the film. Data from three devices is included in this article: Contact 1 and Contact 2 were fabricated on the same heterostructure, while Contact 3 was fabricated on another heterostructure grown under identical conditions.

We used fits with the Blonder–Tinkham–Klapwijk (BTK) model to identify the superconducting contribution of Nb in a 175 mT field where the QAHI is in the trivial insulating state. The BTK theory analyzes the differential conductance vs. bias voltage with the superconducting order parameter $\Delta$ and barrier height Z as fitting parameters. We used a modified BTK model that takes into account the finite quasi-particle lifetime or broadening parameter $\Gamma$ [35]. Our fitting shows smaller $\Gamma$ in Fig. 1d than in Fig. 1e. One reason is that the temperature raises from 15 mK to 320 mK, which increases the broadening effect, and the other reason is that data in Fig 1d was obtained with a much more transparent contact. It has been shown that the less transparent a contact, the larger $\Gamma$ is [35].

$$G_{normalized} = \frac{dI}{dV}(V)$$

$$= 2N(0)ev_F S \int_{-\infty}^{+\infty} \frac{\partial f_0(E-eV)}{\partial(eV)}[1 + A(E) - B(E)]dE$$

$$= 2N(0)ev_F S \int_{-\infty}^{+\infty} \frac{\exp(\frac{E-eV}{kT})}{kT[1+\exp(\frac{E-eV}{kT})]^2}[1 + A(E) - B(E)]dE$$

*A* represents the Andreev reflection possibility, *B* the normal reflection, where:

$A = \frac{\Delta^2}{E^2+(\Delta^2-E^2)(1+2Z^2)^2}$ and $B = 1 - A$. Note that the fitting parameters, such as the tunneling barrier height thus obtained apply only to the Nb contribution, while the QAHI parameters may differ slightly.

**Author contributions**

J.S., J.L. and O.A. prepared the point contact devices and performed all experiments with help of C.-W.C, Y.J.H, K.Y.Y. and S.K.G.; R.L. and K.T.L. planned the experiments. R.L. analyzed the experimental data with help from J.S. and J.L.; L.P, Q.L.H. and K.L.W. provided the QAHI films. Z.J.C.

and K.L. prepared the Nb layer of the heterostructures. Y.-M.X, J.Z.G and C.-Z.C. performed the numerical simulations and K.T.L. provided the theoretical support. R.L. wrote the manuscript with contributions from J.S., Q.L.H., J.Z.G., C.-Z.C., K.L.W. and K.T.L., all authors contributed to the discussion and data interpretation and have read and approved the final manuscript.

**Acknowledgements**
R.L. acknowledges enlightening discussions with X.-G. Wen, H. Zhang, M. He and X. Dai and acknowledges support by grants from the Research Grants Council of the Hong Kong Special Administrative Region, China (GRF-16302018, SBI17SC14, IEG16SC03). K.T.L. acknowledges the support of HKRGC through C6026-16W, the Croucher Foundation and the Tai-chin Lo Foundation. K.L.W. and Q.L.H. acknowledge the support of the ARO Multiple University Research Initiative (MURI) and Energy Frontier Research Center (SHINES). Q.L.H. also acknowledges the supports from the National Natural Science Foundation of China (Grant No. 11874070), the National Key R&D Program of China (Grant 338 No. 2018YFA0305601), and the National Thousand-Young Talents Program in China. Z.J.C. and K.L. acknowledge support from the US NSF (DMR-1610060).

# Spectroscopic Fingerprint of Chiral Majorana Modes at the Edge of a Quantum Anomalous Hall Insulator / Superconductor Heterostructure

J. Shen[1], J. Lyu[1,2], J. Z. Gao[1], Y.-M Xie[1], C.-Z. Chen[1], C.-w. Cho[1], O. Atanov[1], Z. Chen[3,4], K. Liu[3,4], Y. J. Hu[5], K. Y. Yip[5], S. K. Goh[5], Q. L. He[6,7], L. Pan[6], K. L. Wang[6], K. T. Law[1] and R. Lortz[1]

[1]*Department of Physics, The Hong Kong University of Science and Technology, Clear Water Bay, Kowloon, Hong Kong.*
[2]*Department of Physics, Southern University of Science and Technology, 1088 Xueyuan Road, Nanshan District, Shenzhen, Guangdong Province, China*
[3]*Physics Department, University of California, Davis, CA 95616, USA.*
[4]*Physics Department, Georgetown University, Washington, DC 20057, USA.*
[5]*Department of Physics, The Chinese University of Hong Kong, Shatin, New Territories, Hong Kong.*
[6]*Department of Electrical and Computer Engineering, Department of Physics, and Department of Materials Science and Engineering, University of California, Los Angeles, CA 90095, USA.*
[7]*International Center for Quantum Materials, School of Physics, Peking University, Beijing, 100871, China.*

**Section 1: Additional data supporting our interpretation**

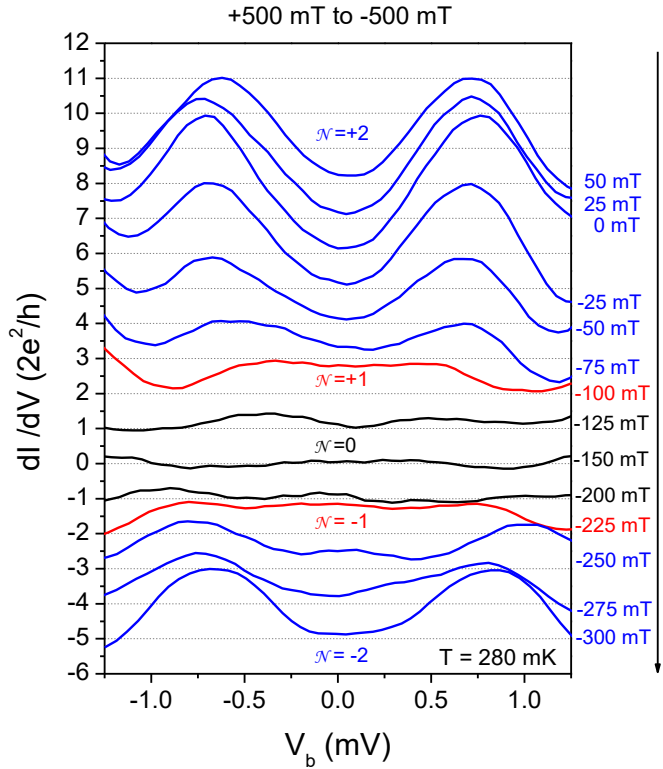

**Fig. S1.** Differential conductance of a point contact (Contact 1) at the edge of the $(Cr_{0.12}Bi_{0.26}Sb_{0.62})_2Te_3$ / superconductor heterostructure at higher temperature of 280 mK. The data was taken starting from a positive field by decreasing the field in 25 mT decrements to negative fields, thus inducing the QAHI's magnetization reversal. Dips surrounded by two coherence-like peaks mark the $\mathcal{N} = \pm2$ topological superconducting region, while a flat plateau at -100 mT is observed, which is attributed to the $\mathcal{N} = 1$ topological superconducting phase shortly before the QAHI enters the trivial insulating regime. When the absolute value of the magnetic field is increased further, a re-entry of the $\mathcal{N} = -1$ phase in the form of a plateau at -225 mT can be detected, and finally the characteristic dip of the $\mathcal{N} = -2$ phase occurs again at zero bias. Offsets of $2e^2/h$ have been added for better clarity, except for the −150 mT data.

In Fig. S1 we show additional data of Contact 1 recorded immediately after measuring the magnetization cycle shown in the main article in Fig. 2. Here we increased the temperature to 280 mK to test the effect of temperature. The data show a very similar trend where dip-like structures surrounded by two peaks (attributed to the $\mathcal{N}=2$ state) change to a plateau-like structure without dip (attributed to the $\mathcal{N}=1$ state) before the conductance vanishes in the field range of -125 mT to -200 mT. In higher fields, the reverse sequence occurs, with the plateau-like feature re-emerging at -225 mT, replaced by a re-entry of the dip-like features associated with the $\mathcal{N}=-2$ state.

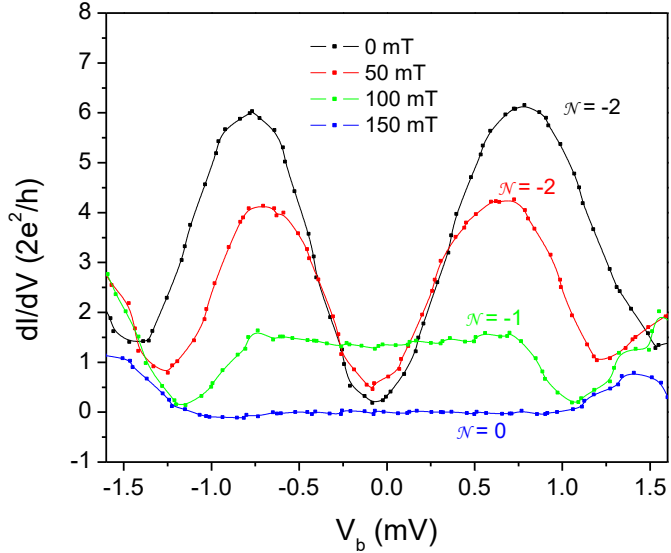

**Fig. S2.** Additional data from Contact 1 after subtraction of the Nb contribution measured at 15 mK directly after the initial cooling of the sample from room temperature. It clearly shows the different regimes that occur before the reversal of the magnetization the QAHI. No shift was applied here to highlight the different d$I$/d$V$ spectra in the 3 different regimes.

In Fig. S2 we show another data set for Contact 1, which was measured immediately after the initial cooling of the sample from room temperature. The Nb background curve was subtracted. The transition from a dip surrounded by two peaks at 0 and 50 mT to a plateau at 100 mT can be seen before the conductance vanishes at 150 mT, illustrating the sequence of the topological transitions from $\mathcal{N}=-2$ over $\mathcal{N}=-1$ to the trivial insulating state.

In Fig. S3 we show additional data of another contact (Contact 3) fabricated on another heterostructure grown under identical conditions (measured at 15 mT). The same features attributed to topological transitions with dip-like features around zero bias ($\pm\mathcal{N}=2$ state) and plateau-like features ($\pm\mathcal{N}=1$ state) can be seen before and after the conductance vanishes in the trivial insulating state between 90 and 100 mT. The hysteresis range of this heterostructure was somewhat smaller, so that the topological transitions occur at slightly lower fields.

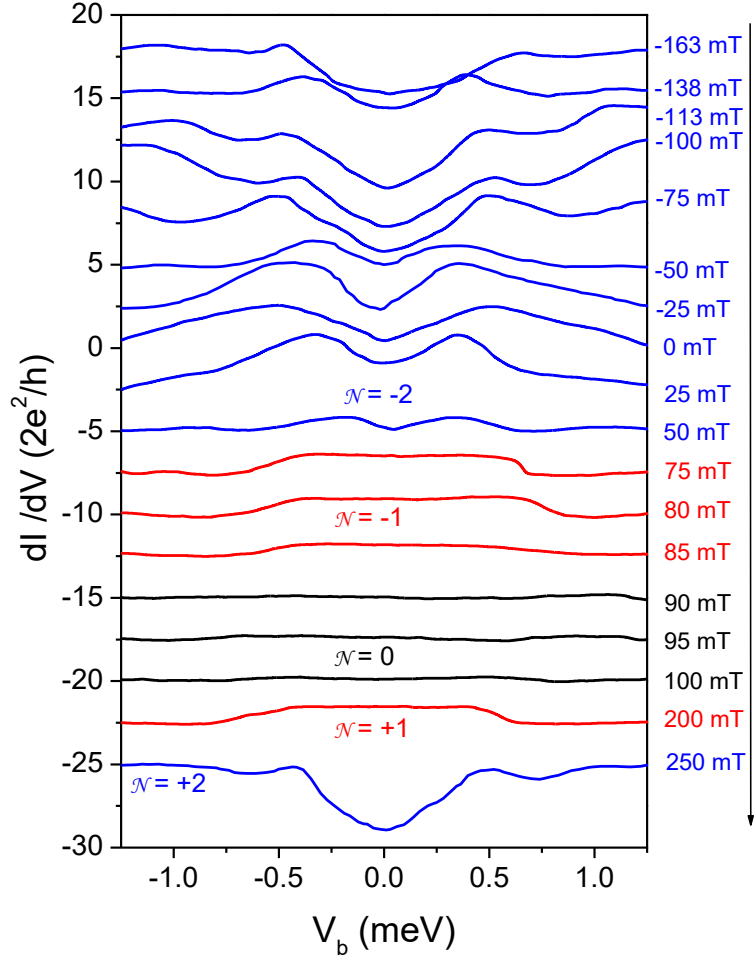

**Fig. S3.** Differential conductance of a point contact (Contact 3) at the edge of another $(Cr_{0.12}Bi_{0.26}Sb_{0.62})_2Te_3$ / superconductor heterostructure at 15 mK. The data was taken starting from a negative field by increasing the field in small increments to positive fields, thus inducing the QAHI's magnetization reversal. Dips surrounded by two coherence-like peaks mark the $\mathcal{N} = \pm 2$ topological superconducting regions, while flat plateaus at +75 mT, +80 mT and +200 mT are observed, which are attributed to the $\mathcal{N} = \pm 1$ topological superconducting phase. The latter occur shortly before and after the trivial insulating regime of the QAHI. Offsets of 2.5 x $2e^2/h$ have been added for better clarity.

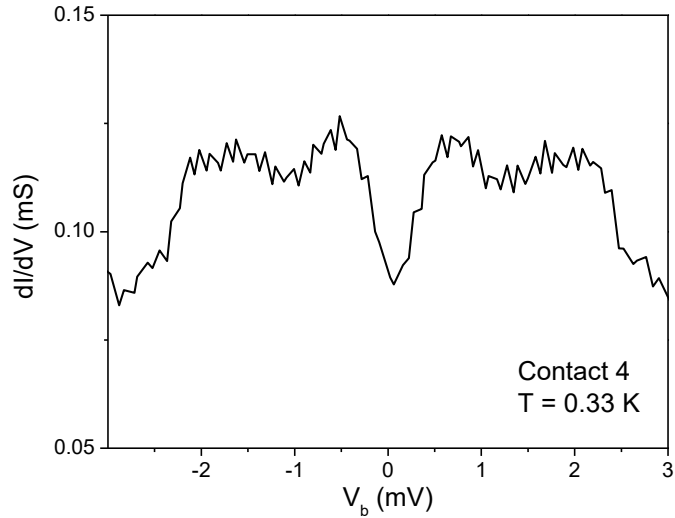

**Fig. S4.** Raw data of the differential conductance of another point contact (Contact 4) at the edge of a $(Cr_{0.12}Bi_{0.26}Sb_{0.62})_2Te_3$ / Nb heterostructure in zero field. The dip at zero bias associated with the $\mathcal{N} = +2$ state is visible here without background subtraction.

In Fig. S4 we show another set of additional raw data of another contact (Contact 4) fabricated on a third heterostructure. The contact had a medium transparency and is dominated by the superconducting QAHI. The characteristic dip around zero bias framed by two peaks associated with the $\mathcal{N} = \pm 2$ topological superconducting region is clearly visible in the raw data. The additional features at higher bias voltage are due to the bulk bands of the QAHI.

## Section 2: Closure of the bulk gap at the topological transitions

The theoretical simulations predict a closure of the bulk gap in the QAHI when the transitions between the different topological regimes occur. Such a closure is not clearly visible in our data. If present, this closure seems to happen quite abruptly, and the exact fields in which it occurs may have been missed in our experiments. Nevertheless, there is some evidence of a significant reduction in the gap before the transition from the $\mathcal{N} = -2$ to the state $\mathcal{N} = -1$ in the data of Contact 2, for which we have measured with smaller field increments. Fig S4 a) shows the relevant data in which we have marked the peak positions of the two peaks that frame the zero bias dip of the $\mathcal{N} = -2$ state. This gives an approximate measure of the variation of the gap size. The distance between these peak pairs, which occur at positive and negative bias voltage, shrinks significantly before the transition to the plateau-like feature. In Fig. S4 b) we record the distance of the peak positions as a function of the applied field. It shows that the size of the gap decreases significantly. A linear extrapolation helps us to estimate the gap closure field as 68 mT for this heterostructure.

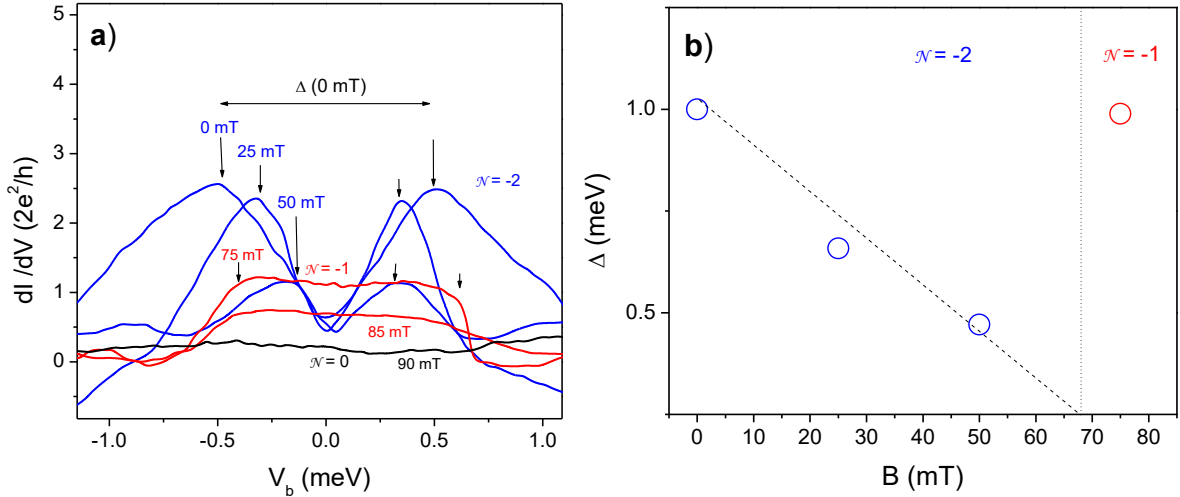

**Fig. S5. (a)** Selected data of Contact 3 after subtraction of the Nb background, here plotted without shift. Arrows mark the positions of the two peaks surrounding the zero-bias dip of the $\mathcal{N} = -2$ state. A decrease of the distance between the two peaks can be clearly seen, which may be an indication for a gap closure between 50 and 100 mT. **(b)** Plot of the distance between the positions marked by the arrows in a) as a function of applied field. The decrease of the peak positions from 0 to 50 mT can be clearly seen. A linear extrapolation provides a characteristic field of 68 mT where the gap closure may occur.

## Section 3: Green's Function Formalism

The d$I$/d$V$ spectra for the QAHI / SC junction can be calculated using the scattering matrix approach

$$G_c(E) = \frac{e^2}{h} Tr[I - R_{ee}(E)R_{ee}^\dagger(E) + R_{he}(E)R_{he}^\dagger(E)]$$

where $R_{ee}(E)$ and $R_{he}(E)$ describe the scattering matrices for normal and Andreev reflections, respectively. The scattering matrices can be evaluated using the recursive Greens function method as follows:

$$R_{\alpha\beta}(E) = -\delta_{\alpha\beta}I + i[\Gamma_\alpha^{\frac{1}{2}}(E)G^R{}_{i,\alpha\beta}E)\Gamma_\beta^{1/2}(E)]$$

where $\alpha,\beta = \{e/h\}$ are the electron/hole indices and $I$ is the identity matrix. Alternatively, in the presence of translational invariance in the $y$ direction,

$$R_{\alpha\beta}(E,k_y) = -\delta_{\alpha\beta}\tilde{I} + i[\Gamma_\alpha^{\frac{1}{2}}(E,k_y)G^R{}_{i,\alpha\beta}(E,k_y)\Gamma_\beta^{1/2}(E,k_y)]$$

for a fixed momentum $k_y$ with the identity matrix $\tilde{I}$. Here $G^R{}_i(E) = [E + i0^+ - \mathcal{H}_{BdG} - \Sigma^R]^{-1}{}_{ii}$ is the retarded Greens function at the interface between the lead and the QAHI / SC junction. $\mathcal{H}_{BdG}$ is the Hamiltonian of the scattering region, and $\Gamma$ is the self-energy of the lead. The $\Gamma$ is defined as $\Gamma_\alpha = i[\Sigma^R{}_\alpha - \Sigma^A{}_\alpha]$ where $\Sigma^A{}_\alpha$ is the retarded / advanced self-energy of the α-particle in the lead. We note that the total differential conductance $G_c(E)$ is related to momentum resolved conductance by summing the traces of all sub-blocks:

$$G_c(E) = \tfrac{e^2}{h}\Sigma_{k_y}Tr[I - R_{ee}(E,k_y)R_{ee}{}^\dagger(E,k_y) + R_{he}(E,k_y)R_{he}{}^\dagger(E,k_y)].$$

**Section 4: The lattice Hamiltonian for the tunneling conductance calculations**

In the simulation, we calculate the d$I$/d$V$ using a lead-QAHI geometry (see Fig. S6). We apply a periodic boundary condition along the $x$-direction so that we can use a good quantum number $k_x$ to label each tunneling channel.

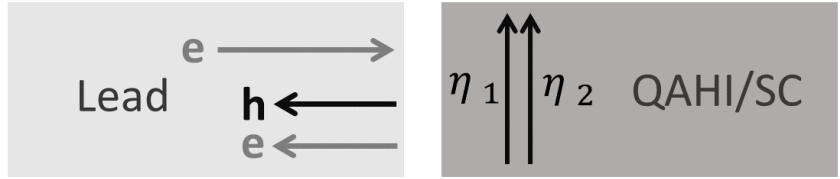

**Fig. S6.** A schematic representation of the simulation geometry. An injected electron interacts with a chiral Majorana mode ($\eta_{1,2}$), and is reflected as electron (e) or hole (h).

The superconducting QAHI sample located in the region $1 \leq n_y \leq N_y$ is described by a lattice Hamiltonian:

$$\mathcal{H}_{TI} = \Sigma_{k_x,1\leq n_y \leq N_y}\ \psi^\dagger_{k_x,n_y}\Big[(m_0 - 4m_1 + 2m_1\cos k_x)\tau_x s_z - \hbar v_F \sin k_x \tau_z \sigma_y s_z +$$
$$M_z\sigma_z s_z + \Delta\tfrac{\tau_z+I}{2}\sigma_y s_y - \mu s_z\Big]\psi_{k_x,n_y} + \psi^\dagger_{k_x,n_y}\Big(m_1 \tau_x s_z - \tfrac{i\hbar v_F}{2}\tau_z\sigma_x\Big)\psi_{k_x,n_y+1} + \text{h.c.}.$$

This Hamiltonian is obtained by discretizing the Hamiltonian $H_{TI}(k)$ in the main text. Here $\psi = [\varphi_{t\uparrow}, \varphi_{t\downarrow}, \varphi_{b\uparrow}, \varphi_{b\downarrow}, \varphi^\dagger_{t\uparrow}, \varphi^\dagger_{t\downarrow}, \varphi^\dagger_{b\uparrow}, \varphi^\dagger_{b\downarrow}]^T$ denoting the Nambu basis, $\varphi^\dagger$ is the electron creation operator for the QAHI, $\tau$, $\sigma$, $s$ operate on layer, spin, particle-hole space, respectively, and $I$ is the unit matrix. We set the parameters $\hbar v_F = 3\,\text{eV}\cdot\text{Å}$, $m_0 = -5\,\text{meV}$, $m_1 = 15\,\text{eV}\cdot\text{Å}^2$, which are estimated from the dispersion of the TI surface states and the experimental data of a TI thin film [24]. In the simulation we set $\mu = 0$, $\Delta = |m_0|$, and the lattice constant $a$ to be 4 nm.

The lattice Hamiltonian $\mathcal{H}_L$ for the semi-infinite lead located at $n_y < 0$, is given by

$$\mathcal{H}_L = \sum_{k_x, n_y < 0} \phi^\dagger_{k_x, n_y} (4t_L - \mu_L - 2t_L \cos k_x) s_z \phi_{k_x, n_y} - t_L \phi^\dagger_{k_x, n_y} s_z \phi_{k_x, n_y+1} + \text{h.c.}.$$

Here $\phi = [c_\uparrow, c_\downarrow, c_\uparrow^\dagger, c_\downarrow^\dagger]^T$, $c^\dagger$ is the electron creation operator within the lead. We have chosen the hopping of the lead to be $t_L = \hbar v_F$ and its chemical potential $\mu_L = 2t_L$.

The coupling between lead and sample is given by:

$$\mathcal{H}_c = \sum_{k_x} -t_c \psi^\dagger_{k_x, n_y=1} s_z \phi_{k_x, n_y=0} + \text{h.c.}.$$

The transparency of the tunneling junction is characterized by the amplitude of $t_c$. For the sake of simplicity, we have assumed that the coupling strength $t_c$ is $k_x$ independent.

The total Hamiltonian is written as

$$\mathcal{H} = \mathcal{H}_{TI} + \mathcal{H}_L + \mathcal{H}_c.$$

With this Hamiltonian we can evaluate the tunneling conductance using the Green's function formalism. In the simulation performed for Fig. 3 (b) and Fig. 4 in the main text, $t_c = 0.1$ and $t_L = 0.1\hbar v_F$.

**Section 5: Comments on the height of the plateau-like feature in d$I$/d$V$ in the $\mathcal{N} = \pm\,1$ phase**

In a finite QAHI system, the chiral Majorana edge modes have discretised momentum, meaning that the edge modes are separated by a small energy difference. If the resolution of the tunnelling probe were high enough (e.g. in the weak coupling limit and at temperatures where $k_BT$ is much lower than the energy spacing of the modes), one would observe many tunnelling peaks in the energy interval corresponding to the superconducting gap due to the Majorana edge modes, as shown in Fig. S7 for the the $\mathcal{N} = \pm\,1$ phase. In Fig. S7, $t_c$ is the parameter that characterizes the coupling strength between the lead and the superconductor in our theoretical model, as explained in Section 4 above. It can be shown that the Andreev reflection probability is proportional to $t_c^2$ and thus related to the effective tunnelling barrier height in the BTK model. With finite thermal broadening, however, the peaks are washed out and merge in the $\mathcal{N} = \pm\,1$ phase to form a flat plateau. For a weak coupling junction with $t_c = 0.05$ the plateau height is less than $2e^2/h$. However, if the coupling is increased, which corresponds to a more transparent junction, the plateau approaches the quantised value of $2e^2/h$ for $t_c = 0.1$. If the coupling strength is increased beyond this value, the plateau may be higher. Therefore, the height of the plateau is not universal for such dispersive chiral Majorana modes.

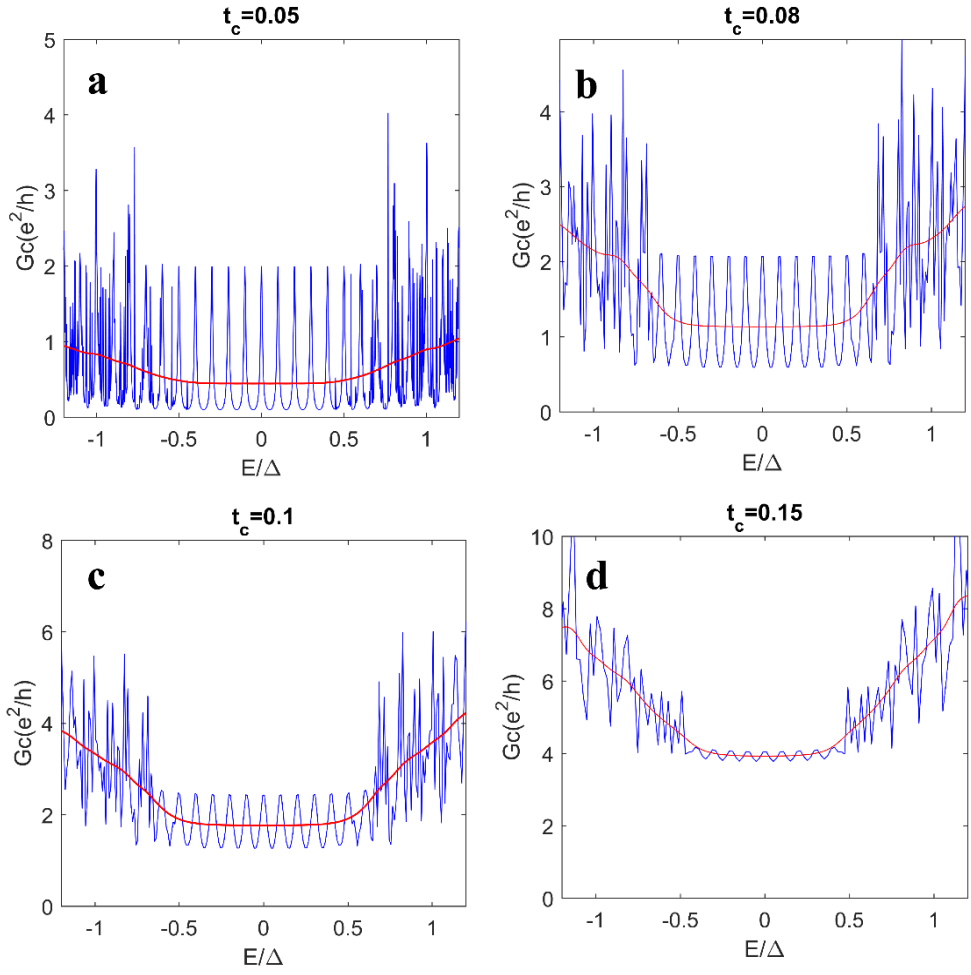

**Fig. S7.** Simulated conductance of a finite system vs. energy in units of the superconducting gap Δ for different transparencies $t_c$ (in units of $\hbar v_F$) of the tunneling contact. The blue data simulate a conductance with a high energy resolution of the tunneling probe and thus resolve a series of discrete Majorana peaks corresponding to different momentum and energy of the chiral Majorana edge modes. The red curves take into account a finite thermal broadening, whereby the peaks are washed out to form a constant plateau-like feature within the energy range of the superconducting gap Δ.